\documentclass[pdflatex,sn-mathphys]{sn-jnl}

\usepackage{mathtools}
\usepackage{hyperref}
\usepackage{rotating}

\jyear{2022}%
\begin{document}

\title[A New Parameterized Interacting Holographic Dark Energy]{A New Parameterized Interacting Holographic Dark Energy}


\author*[1]{\fnm{Celia} \sur{Escamilla-Rivera}}\email{celia.escamilla@nucleares.unam.mx}

\author*[1]{\fnm{Aldo} \sur{Gamboa}}\email{aldojavier@ciencias.unam.mx}

\affil[1]{\orgdiv{Instituto de Ciencias Nucleares}, \orgname{Universidad Nacional
	Aut\'onoma de M\'exico}, \orgaddress{\street{Circuito Exterior C.U.}, \postcode{04510}, \state{M\'exico D.F.}, \country{M\'exico}}}


\abstract{We study a cosmological model based on the holographic principle that allows an interaction between dark energy and dark matter with a Hubble infrared cutoff. We adopt an agnostic point of view with respect to the form of the interaction between these dark components by using redshift parameterizations of the dark energy equation of state (DE EoS). Following this approach, we obtain an analytical expression of the Hubble parameter and we test it with four different parameterizations of the DE EoS using currently Pantheon supernovae, Cosmic Chronometers and Galaxy Clustering data sets. Our analyses show that the considered parameterizations on this agnostic proposal allow a late accelerated cosmic expansion in comparison to the results from previous works where specific forms of the interaction are employed but fail to explain this behaviour. Furthermore, we find that \textit{our proposal solves the so-called coincidence problem in Cosmology and in this direction shed some light on the problem of the Hubble tension.}}



\maketitle


\section{Introduction}
\label{sec:introduction}

The discovery of the accelerated expansion of the Universe \cite{SupernovaSearchTeam:1998fmf, SupernovaCosmologyProject:1998vns} gave rise to one of the most fundamental problems in physics: the nature of Dark Energy (DE). The standard pivot model, the concordance model $\Lambda$CDM, introduces a cosmological constant as the source of this dark component, whose equation of state (EoS) is $w_{\text{de}} = p_{\text{de}}/\rho_{\text{de}}=-1$, where $p_{\text{de}}$ and $\rho_{\text{de}}$ are the pressure and energy density of DE, respectively, reproducing the observed repulsive force needed to explain the accelerated cosmic expansion. So far, this \textit{vanilla} model has the best performance fitting presently observational data \cite{Perivolaropoulos:2021jda}. However, the nature of one of the main characters of this concordance model, the cosmological constant $\Lambda$, has not been clarified yet. In particular, if we attempt to identify the DE density with the predicted quantum vacuum energy, then a difference in several orders of magnitude arises between the inferred and predicted values \cite{weinberg,carroll,Rugh:2000ji,Stenflo:2020zsc}. Additionally, within the $\Lambda$CDM model, the DE density $\rho_{\text{de},0}$ and Dark Matter (DM) density $\rho_{\text{dm},0}$ at the present epoch $z=0$, are of the same order of magnitude, $\rho_{\text{de},0}/\rho_{\text{dm},0} \sim \mathcal{O}(1)$, which seems to indicate that we are living in a special period of the Universe since these densities evolve differently over time. Nevertheless, this fact poses a problem because it requires very specific initial conditions in the early Universe. This is the so-called  \textit{coincidence problem} \cite{Velten:2014nra,Cunillera:2021izz}. Our uncertainty about the nature of the cosmological constant and the associated theoretical problems have motivated an extensive study of alternative DE models \cite{Yang:2021hxg,Schoneberg:2021qvd,DiValentino:2019ffd}. 

A particularly interesting model which couples the features of DE with quantum gravity is the \textit{holographic dark energy} (HDE). As the name suggests, this model is based on the \textit{holographic principle} (HP), which is an expected property of any quantum gravity theory \cite{hp_intro}. This HDE model has been successfully applied to several fields of physics, e.g in nuclear physics \cite{Liu_2007}, condensed matter \cite{Hartnoll:2009sz} and cosmology \cite{Strominger:2001pn}. According to the HP, the number of degrees of freedom in a physical system scales with the area of its boundary \cite{bekenstein, tHooft:1993dmi, Susskind:1994vu, hp}. In this context, in \cite{Cohen_hde} was suggested that, for an effective quantum field theory in a box of size $L$, its ultra-violet cutoff is related to the infrared (IR) cutoff  due to a limit set by the formation of a black hole, which results in an upper bound to the zero-point energy density of the box. Motivated by the cosmological constant problem, in \cite{original_hde} these results were applied to the Universe to obtain an expression for the DE density which saturates the mentioned upper limit of avoiding the collapse into a black hole. The resulting expression is
\begin{equation}\label{eq:rho PH}
	\rho_{\text{de}} =  3 C^2 M_{\text{Pl}}^2 L^{-2},
\end{equation}
where $C$ is a constant, $M_{\text{Pl}} = \sqrt{1/8 \pi} $ is the reduced Planck mass\footnote{In this paper we will use geometrical units so that $c=1$ and $G=1$, unless otherwise specified.} and $L$ is the IR cutoff which represents a characteristic length scale of the Universe. The problem of DE now reduces to find $C$ and to choose an appropriate IR cutoff.
The first natural choice for $L$ is the inverse of the Hubble parameter, $L = H^{-1}$ (the so-called \textit{Hubble IR cutoff}). Nevertheless, this choice implies that $w_{\text{de}} = 0$, which is an EoS that does not allow an accelerated expansion of the Universe within the HDE model \cite{HSU200413, original_hde}. Another choice for the length scale is the \textit{future event horizon} \cite{original_hde}, $L = a \int_{t'}^\infty dt' / a(t')$, where $a$ is the scale factor and $t'$ is the cosmic time. With this choice, the value of the constant $C$ in \eqref{eq:rho PH}, must be estimated with observations. Overall, depending on the choice of the IR cutoff, the HDE proposal gives a very competitive cosmological model. For example, within this framework, the presence of curvature in the Universe has been analysed \cite{Huang_2004}, different bounds for the neutrinos masses have been estimated \cite{zhang_neutrino, wang_neutrino}, the inflationary period has been studied \cite{CHEN2007256} and also several modified theories of gravity have been explored \cite{review_hde}. 

Another issue analysed using the HDE model is the \textit{Hubble constant tension}, an important problem in cosmology that arises due to the discrepancy between the estimated values of the Hubble constant, $H_0$, at the early and late Universe epochs \cite{DiValentino:2020zio, Riess:2021jrx}. The measurements of the cosmic microwave background (CMB) radiation from Planck satellite \cite{Planck:2018vyg} give the value $ H_0 = 67.4 \pm 0.5 {\rm \,km\, s^{-1}\, Mpc^{-1}}$, whereas a combination of different local measurements \cite{Verde:2019ivm} yields a representative value of $ H_0 = 73.1 \pm 0.9 {\rm \,km\, s^{-1}\, Mpc^{-1}}$, which is $5.7 \sigma$ discrepant from the Planck result. Recently, the Pantheon collaboration has reanalyzed the calibration used for the photometry of the supernovae type Ia sample, including the calibration used to obtain the SH0ES results~\cite{Brout:2021mpj, Scolnic:2021amr}, adding a new value from Pantheon$+$Cepheid-Supernovae Ia baseline $H_0 = 73.04\pm 1.04{\rm \,km\, s^{-1}\, Mpc^{-1}} $ \cite{Riess:2021jrx}. The contribution to the Hubble constant from the systematic supernovae calibration is smaller than $\sim 0.2{\rm \,km\, s^{-1}\, Mpc^{-1}}$, so that the tension cannot be attributed to this type of systematics.

Several mechanisms and explanations have been proposed in order to alleviate the Hubble tension \cite{DiValentino:2021izs,DiValentino:2020zio}. In particular, in \cite{h0_HDE} it was shown that, within the HDE model with the future event horizon as the IR cutoff, and depending on the value of $C$, the EoS of HDE satisfies $w_{\text{de}} > -1$ at an early epoch ($z \gtrsim 1$), and $w_{\text{de}} < -1$ at a later epoch ($z \lesssim 1$). Therefore, the Universe has a smaller acceleration earlier on and faster acceleration afterwards. This \textit{delayed acceleration} successfully alleviates the Hubble tension because the value of $H_0$ estimated from local measurements is higher than the value calculated with the observations of the CMB. 

Another fundamental problem in cosmology is the nature of DM \cite{dm_review, Salucci:2018hqu}. There are several cosmological and astrophysical scenarios whose theoretical explanations require the presence of this exotic entity in order to fit the observational data. For example, the measurements of galaxy rotation curves \cite{Brownstein_2006}, the dispersion velocities of galaxies in clusters \cite{Faltenbacher:2006rb}, the presence of gravitational lensing \cite{Massey:2010hh}, large scale structure formation \cite{Primack:1997av} and the distribution of anisotropies in the CMB \cite{Planck:2018vyg}. However, despite the big theoretical and experimental efforts, we still do not arrive to a unique agreement on the physics of DM, and its unknown nature has led us to explore several possible candidates and exotic models \cite{Bergstrom:2009ib}.

Among these models, exists the possibility that DM is coupled to DE. These proposals are generically known as interacting DE (IDE) models \cite{Wang_2016, DiValentino:2019ffd}. Usually, this coupling is realised through an interaction term $Q$ included in the continuity equations of DE and DM [see Eqs.~\eqref{eq:cont_eq rho_de} and \eqref{eq:cont_eq rho_dm}]. Several specific forms of $Q$ have been considered in the literature, but in general it is assumed to be a function of $\rho_{\text{de}}$ and $\rho_{\text{dm}}$. The three most widely used options are $Q = \Gamma_1 H \rho_{\text{de}}$, $Q = \Gamma_2 H \rho_{\text{dm}}$ and $Q = \Gamma_1 H \rho_{\text{de}} + \Gamma_2 H\rho_{\text{dm}}$, where $H$ is the Hubble parameter and $\Gamma_1$, $\Gamma_2$ are constants, although there have been considered more intricate forms of the interaction \cite{delCampoo,Wang_2016}. However, an important theoretical caveat to these models is that the interaction term is introduced in a phenomenological way, i.e. it is not directly derived from an action, making the choice of the form of $Q$ somewhat arbitrary.
The coupling of DE and DM has already been studied in the context of the HDE \cite{Wang:2005jx, PAVON2005206, Ma:2007pd, Zadeh:2016vgc,  delcampo}, however, the specific couplings in these interacting HDE (IHDE) models do not come from an action principle either. In particular, depending on the choice of the interaction term and the IR cutoff, we could have a universe with \cite{Nayak:2019njd} or without \cite{Zadeh:2016vgc} accelerated expansion. Therefore, a more general framework is needed in order to address the viability of the IHDE model without depending on the explicit form of the interaction.

In this paper, we present a generic treatment of the coupling between DE and DM in the IHDE model with a Hubble IR cutoff. As we mentioned earlier, if there is no coupling with DM, then this cutoff choice is discarded because it yields a wrong DE EoS. As we will show, the interaction allows us to reconsider this IR cutoff because it predicts an EoS which indeed allows an accelerated expansion. This treatment is achieved by quantifying the interaction $Q$ through a DE EoS redshift parameterization, instead of choosing a specific coupling of the dark sector. We demonstrate the viability of this parameterized IHDE model by doing statistical analyses with observational data for various DE EoS parameterizations.

This paper is organised as follows: in Sec.~\ref{sec:ihde}, we present the cosmological equations used for the parameterized IHDE scenario and we use them with the Hubble IR cutoff to obtain an analytical expression for the Hubble parameter. In Sec.~\ref{sec:dark universe}, we use the data from certain cosmological observations and various DE EoS parameterizations to study the consequences of the IHDE model in a Universe dominated by matter and DE. Finally, in Sec.~\ref{sec:conclusions} we present our conclusions and suggestions for future work.


\section{A New Interacting Holographic Dark Energy}
\label{sec:ihde}

According to the IHDE model, the cosmological evolution is determined by:
\begin{enumerate}
    \item the DE density obtained from the HP with a certain choice of the IR cutoff,
\item the standard Friedmann equation, 
\item and the coupling term for DM and DE densities. 
\end{enumerate} 
Combining these three ingredients, we can obtain an analytical or numerical solution of the Hubble parameter, which will allow us to calculate 
$H_0$ with standard parameter statistics estimation techniques.

We start with the Friedmann equation, 
\begin{equation}\label{eq:friedmann}
3 M_{\text{Pl}}^2 H^2 = \sum_j \rho_j,
\end{equation}
written in terms of the Hubble parameter $H\coloneqq\dot a / a$, where $a$ is the scale factor and the dot denote derivatives with respect to cosmic time $t$, and the energy densities of radiation, baryonic matter, cold DM, DE and curvature $k$, denoted collectively as $\rho_j$, where $j = \{\text{r}, \text{b}, \text{dm}, \text{de}, k \}$, respectively, with $\rho_k = - 3 M_{\text{Pl}}^2 k / a^2$.

The radiation and baryon densities have the standard continuity equations. In contrast, the modified version of the continuity equations for DE and DM are 
\begin{gather}
	\dot\rho_{\text{de}} + 3 H \rho_{\text{de}} (1 + w_{\text{de}}) = Q  , \label{eq:cont_eq rho_de}\\
	\dot\rho_{\text{dm}} + 3 H \rho_{\text{dm}}  =  -Q, \label{eq:cont_eq rho_dm}
\end{gather}
where $Q$ is the interaction term which allows an energy transfer between the two dark components. The sign of $Q$ is chosen so that if $Q>0$ then DM is transferring energy to DE, and otherwise for $Q<0$. 

Differentiating Eq.~\eqref{eq:friedmann} with respect to the redshift $z = a^{-1} - 1$ (so that $ \frac{d}{dt} = - (1+z)H \frac{d}{dz} $) we get an equation relating the derivatives of the energy densities. Therefore, substituting Eqs.~\eqref{eq:cont_eq rho_de} and \eqref{eq:cont_eq rho_dm}, and the standard continuity equations for all the other components, we obtain the evolution of the Hubble parameter as
\begin{equation}\label{eq:prime H}
	H'(z) =  \frac{3 H}{2(1+z)} + \frac{\rho_{\text{r}} - \rho_k + 3 \rho_{\text{de}}w_{\text{de}}}{ 6 M_{\text{Pl}}^2 (1+z) H }   ,
\end{equation}
where the prime denotes derivative with respect to $z$. Notice that the interaction between DE and DM is implicitly contained in the EoS $w_{\text{de}}$ [see Eq.~\eqref{eq:Q_LisH}].

We will choose the Hubble horizon as the IR cutoff, $L= H^{-1}$. We start evaluating Eq.~\eqref{eq:rho PH} at present time $t_0$. Using the definition of the fractional densities  $\Omega_{j} \coloneqq \rho_j / 3 H^2 M_{\text{Pl}}^2$, we find
\begin{equation}
 C^2 = \Omega_{\text{de},0} =\Omega_{\text{de}}, 
\end{equation}
where the subscript $0$ denotes evaluation at time $t_0$. Therefore, with the Hubble IR cutoff, the value of $C$ comes directly from the theory (unlike the future event horizon IR cutoff which requires that $C$ must be inferred from cosmological observations). Remarkably, the fractional DE density will remain the same during the evolution of the Universe. Next, from Eqs.~\eqref{eq:rho PH} and \eqref{eq:friedmann}, we have
\begin{equation}\label{eq:rhos}
\sum_j\rho_j =  \frac{\rho_{\text{de}}}{\Omega_{\text{de},0}} ,
\end{equation}
so that 
\begin{equation}\label{eq:coincidence}
 	\frac{\Omega_{\text{r}} + \Omega_{\text{b}} +  \Omega_{\text{dm}} + \Omega_k }{\Omega_{\text{de}}} = \frac{1- \Omega_{\text{de},0}}{\Omega_{\text{de},0}} = \text{constant} . 
\end{equation}
Notice that \textit{the coincidence problem is solved because the fractional densities in this model must satisfy a constant constriction}. In particular, for a DE and DM dominated Universe, we have $\Omega_{\text{dm}} / \Omega_{\text{de}} \approx \text{constant}$. The fact that the ratio of DM and DE densities is a constant in an IHDE model with $L=H^{-1}$ was previously identified in \cite{PAVON2005206, delcampo}, nevertheless in our work we unambiguously identify this constant and generalise it to include other energy components.

To obtain an expression that relates the interaction term $Q$ with the DE EoS $w_{\text{de}}$, we differentiate Eq.~\eqref{eq:rhos} and substitute the continuity equations for all the components. After a straightforward calculation we find
\begin{equation}\label{eq:Q_LisH}
    \frac{Q}{H} = 3\rho_{\text{de}} w_{\text{de}}(1-\Omega_{\text{de},0}) +   (\rho_ k - \rho_{\text{r}}) \Omega_{\text{de},0}.
\end{equation}
Notice that this expression can be rewritten as $Q \propto \Gamma \rho_{\text{de}}$, (with $\Gamma$ an interaction rate with units of the Hubble constant; in our case $\Gamma = H$), which has been the specific form of the interaction in some previous works \cite{delcampo, Ma:2007pd, review_hde, PAVON2005206, Nayak:2019njd, Wang_2016}. Furthermore, if $Q=0$, then from Eq.~\eqref{eq:Q_LisH} we obtain
\begin{equation}
    w_{\text{de}} = \frac{\rho_{\text{r}} - \rho_k}{3 \rho_{\text{de}}} \frac{\Omega_{\text{de},0}}{1- \Omega_{\text{de},0}}.
\end{equation}
In this latter expression we recognise the known fact  that a non-interacting dark sector ($Q=0$) in a flat late Universe yields an incorrect DE EoS  ($w_{\text{de}} = 0$) when the HP is applied with $L=H^{-1}$ \cite{original_hde}. Therefore, the presence of $Q$ allow us to reconsider this case.

Using Eq.~\eqref{eq:rho PH}, the evolution equation for the Hubble parameter \eqref{eq:prime H} changes as
\begin{equation}\label{eq:evolution H prime}
    H'(z) = \frac{3 H}{2} \frac{1 + \Omega_{\text{de},0} \, w_{\text{de}}}{1+z} + \frac{\rho_\text{r} - \rho_k}{6 M_{\text{Pl}}^2 (1+z) H}. 
\end{equation}
To overcome the lack of enlightenment about the exact form of the interaction term (or, equivalently, the form of $w_{\text{de}}$), we parameterize the DE EoS in terms of the redshift $z$ to account for the effects of $Q$. Then, multiplying Eq.~\eqref{eq:evolution H prime} by $H$, we obtain a first order linear ordinary differential equation for $H^2$. The corresponding analytical solution is
\begin{equation}\label{eq:H_sol_HisL}
\frac{H(z)}{H_0} = \alpha(z)^{-\frac1 2} \sqrt{1 + \int_0^z \alpha(z')\left[ \Omega_{\text{r},0} (1+z')^3 + \frac{k (1+z')}{H_0^2} \right] dz'},
\end{equation}
where we have used that $\Omega_\text{r} = \Omega_{\text{r},0} (1+z)^4$, $\rho_k = - 3 M_{\text{Pl}}^2 k (1+z)^2$ and we have defined
\begin{equation}
    \alpha(z) \coloneqq \exp\left( - 3 \int_0^z \frac{1 + \Omega_{\text{de},0} \, w_{\text{de}}(z')}{1+z'}\, dz'\right).
\end{equation}
Therefore, with Eq.~\eqref{eq:H_sol_HisL} and a given  parameterization of the DE EoS, we can calculate the Hubble parameter for the IHDE model with a Hubble IR cutoff in the presence of radiation, baryonic matter, CDM, DE and curvature.

On the other hand, if we choose to parameterize directly the interaction term $Q = Q(z)$, then one must solve Eq.~\eqref{eq:Q_LisH} for $w_{\text{de}}$ in terms of $Q$, and then substitute back into Eq.~\eqref{eq:evolution H prime}. Doing this, we obtain the differential equation
\begin{equation}
\label{eq:hubble_Qpar}
    \frac{1}{3}\left(H^3\right)' = \frac{3 H^3}{2 (1+z)} + \frac{Q(z) + H(\rho_\text r - \rho_k)}{6 M_{\text{Pl}}^2 (1+z)(1-\Omega_{\text{de},0})}, 
\end{equation}
which does not have an analytical solution unless we consider $\rho_\text r = \rho_k = 0$. However, since it is common to parameterize the DE EoS, we will use Eq.~\eqref{eq:H_sol_HisL} throughout this work.

Finally, as a special case, the $\Lambda$CDM limit can be recovered by equating the expression in \eqref{eq:H_sol_HisL} with the Hubble parameter $H_\Lambda$ of the $\Lambda$CDM model, which is given by
\begin{equation}
\label{eq:H_LCDM}
\frac{H_\Lambda(z)}{H_0} \coloneqq E_\Lambda(z) = \sqrt{ \Omega_{\text{r},0}(1+z)^4 + \Omega_{\text{m},0}(1+z)^3 + \Omega_{k,0}(1+z)^2 + \Omega_{\text{de},0}} \,,
\end{equation}
where $\Omega_{\text{m},0} \coloneqq \Omega_{\text{b},0} + \Omega_{\text{dm},0}$. Performing this, we obtain an expression for $w_{\text{de}}$, which we denote as $w_{\text{de}, \Lambda}$, and is given by
\begin{equation}\label{eq:EoS_LCDM}
    w_{\text{de}, \Lambda}(z) \coloneqq - \frac{1}{ E_\Lambda(z) ^2}.
\end{equation}
Thus, by construction, this EoS give us the Hubble parameter of $\Lambda$CDM model within the framework of IHDE with $L=H^{-1}$. 


\section{Observational test on the New Interacting Holographic Dark Energy}
\label{sec:dark universe}

In order to test the cosmological viability of this parameterized model of DE, we will use current late time observations of the Universe to see if this model can reproduce an accelerated expansion. As we mentioned, the Hubble IR cutoff was discarded in the non-interacting case since it fails to reproduce an accelerated expansion. Nevertheless, a parameterized interacting dark sector allows us to reconsider the simple Hubble IR cutoff since it yields an optimal data fitting behaviour.

For the following analysis, we will assume a flat Universe dominated by matter (baryonic + CDM) and DE. Thus, $k=0$ and $\Omega_{\text r, 0} = 0$, and Eq.~\eqref{eq:H_sol_HisL} takes the form 
\begin{equation}\label{eq:H}
    \frac{H(z)}{H_0} =  \alpha(z)^{-\frac{1}{2}} =  \exp \left( \frac{3}{2} \int_0^z  \frac{1 + \Omega_{\text{de},0} \, w_{\text{de}}(z')}{1+z'} dz' \right).
\end{equation}
Moreover, from Eq.~(\ref{eq:Q_LisH}), we obtain an expression for the interaction term
\begin{equation}
\label{eq:interaction term dark dominated}
Q = 9 M_{\text{Pl}}^2 H^3 w_{\text{de}}(z) \, \Omega_{\text{m},0} \Omega_{\text{de},0},
\end{equation}
where $\Omega_{\text{m},0} = \Omega_{\text{b},0} + \Omega_{\text{dm},0} = 1 - \Omega_{\text{de},0}$.

\subsection{Data analysis}

For the statistical analysis, we perform a Markov chain Monte Carlo (MCMC) Bayesian inference procedure with a modification of the Python implementation \textit{emcee} \cite{emcee}. The likelihood function of the parameter set $\theta$ and their best-fit values, will be determined with the $\chi^2$ statistics.

Observations indicate that the DE domination of the Universe and its accelerated expansion is a recent phenomenon in a cosmological time scale \cite{Sami:2013ssa}. Therefore, if we want to test the capabilities of the parameterized IHDE model to reproduce this behaviour, then we must use low redshift data. In this way, we will use  supernovae Type Ia (SNe Ia) data as standarisable candles (objects whose intrinsic luminosity is well determined), and observational Hubble data (OHD) composed by cosmic chronometers (CC) and baryon acoustic oscillations (BAO)  measurements as standard rulers (objects whose comoving size is well determined). In what follows, we describe our database employed.


\subsubsection{Pantheon SNe Ia data set}

We use the Pantheon SNe Ia compilation \cite{Pan-STARRS1:2017jku} which contains 1048 data points distributed over the redshift interval $0.01 < z < 2.26$. 

To perform the statistical analysis of the Pantheon data set, we employ the theoretical expression of the distance modulus
\begin{equation}\label{eq:modulus_distance}
    \mu_{\text{th}}(z) = 5 \log_{10} \left( \frac{(1+z) c}{h }  \int_0^z \frac{1}{E(z')} \, dz'   \right) + \mu_0,
\end{equation}
where $c$ is the speed of light, $E(z) \equiv H(z)/H_0$ is the normalized Hubble parameter and $h \equiv H_0 / (100 \text{ km s}^{-1}\,\text{Mpc}^{-1})$ is the reduced Hubble constant. For our analysis, we allow the nuisance parameter of the distance estimator $\mu_0$ be fixed according to the prior $H_0 = 73.24 \pm 1.74 \text{ km s}^{-1}\,\text{Mpc}^{-1}$  \cite{Riess:2016jrr} and the cosmological density matter $\Omega_{\text{m}}$ to vary.

The $\chi^2$ function for the best-fit parameter estimation is
\begin{equation}
\chi _{\mathrm{Pantheon}}^{2}(\mathbf{\theta })=\sum_{i=1}^{1048}\frac{(\mu
_{\text{obs}}(z_{i})-\mu _{\text{th}}(z_{i}))^{2}}{\sigma _{\mu,i}^{2}},  \label{eq:chisq}
\end{equation}
where $\sigma_{\mu,i}$ are the associated measurement errors, $\mu
_{\text{obs}}$ is the observed distance modulus and $\mu
_{\text{th}}$ is given by Eq.~\eqref{eq:modulus_distance}.


\subsubsection{Observational Hubble data set}

We employ a compilation of 31 CC $H(z)$ data points \cite{Stern_2010,Moresco_2012, Moresco_2016,10.1093/mnrasl/slv037,Zhang_2014,10.1093/mnras/stx301} measured with the differential age method proposed in \cite{Jimenez_2002}, and 20 galaxy clustering (GC) $H(z)$ data points obtained from baryon acoustic oscillations (BAO) measurements under a $\Lambda$CDM prior \cite{bao1,bao2,bao3,bao4,bao5,bao6,bao7,bao8,bao9,bao10}. The entire OHD sample (CC+GC) consists of 51 data points and it is distributed over the redshift interval  $0.07 < z < 2.36$, and it is shown explicitly in \cite{CC_GC}. We also follow the assumption of \cite{CC_GC} that the BAO data is constructed with independent measurements (not correlated). 

The $\chi^2$ for the OHD  is given by
\begin{equation}
\chi _{\mathrm{OHD}}^{2}(\mathbf{\theta })=\sum_{i=1}^{51}\frac{(H
_{\text{obs}}(z_{i})- H _{\text{th}}(z_{i}))^{2}}{\sigma _{H,i}^{2}},  \label{eq:chisq_H}
\end{equation}
where $\sigma_{H,i}$ are the associated measurement errors, $H_{\text{obs}}$ is the measured Hubble parameter and $H_{\text{th}}$ is the theoretical Hubble parameter of each considered model. 


\subsection{Statistical Analysis}

In the analysis of the parameterized IHDE model, we employ the sets ``Pantheon'' and ``Pantheon+OHD'', whose $\chi^2$ functions are given by
\begin{eqnarray}
\chi^2 = \left\{%
\begin{array}{ll}
\chi _{\mathrm{Pantheon}}^{2},  &  \\
\chi _{\mathrm{Pantheon}}^{2} +  \chi _{\mathrm{OHD}}^{2}. &
\end{array}%
\right.
\end{eqnarray}
Furthermore, we use the gaussian log-likelihood function for the data set $\mathbf{d}$ with measurement errors $\sigma_\mathbf{d}$ given the parameter set $\theta$, 
\begin{equation}
\log {\mathcal{L}}(\mathbf{d \mid \theta })= - \frac{1}{2} \chi^2 + \log(\sigma_\mathbf{d}),
\end{equation}
whose maximization gives the values of the best-fit parameters. 

The different models for the statistical analysis are constructed with the following DE EoS (see Table~\ref{tab:EoS}):
\begin{itemize}
    \item $w_0$: a constant EoS.
    \item $w_{\text{CPL}}$: Chevallier-Polarski-Linder \cite{Chevallier:2000qy, linder}.
    \item $w_{\text{BA}}$: Barboza-Alcaniz \cite{Barboza:2008rh}.
    \item $w_{\Lambda}$: an EoS to measure deviations from $\Lambda$CDM within the IHDE model, inspired by Eq.~\eqref{eq:EoS_LCDM}.
\end{itemize}

The constant parameterization is used as a first \textit{ansatz} due to its simplicity. The CPL and BA parameterizations are used because their two parameters contain information about the present value of the DE EoS and its overall time evolution \cite{Escamilla-Rivera:2016qwv}. The associated Hubble parameters of each DE EoS are calculated with Eq.~\eqref{eq:H} and are also shown in Table~\ref{tab:EoS}. 

\begin{table}[!htbp]
\centering
\begin{tabular}{c|c}
\hline
Equation of state $w_{\text{de}}$ & Dimensionless Hubble parameter $H(z)/H_0$ \\  \hline 
 $w_{0}$ &  $(1+z)^{\frac{3}{2} (1 + \Omega_{\text{de},0} w_0)}$ \\
 $ w_{\text{CPL}} = w_0 + \frac{z}{1+z} w_1$ &  $(1 + z)^{ \frac{3}{2} [1 +  \Omega_{\text{de},0} (w_0 + w_1)]}  \exp\left[-\frac{3}{2}  \frac{\Omega_{\text{de},0} \, w_1 \, z}{1 + z} \right]$ \\
 $w_{\text{BA}} = w_0 + \frac{z(1+z)}{1+z^2} w_1$ &  $(1+z)^{\frac32 (1+\Omega_{\text{de},0} w_0)} (1+z^2)^{\frac34 \Omega_{\text{de},0} w_1}$ \\
 $w_\Lambda = \frac{w_0 - w_1}{  \Omega_{\text m,0}(1+z)^3  + \Omega_{\text{de},0}} + w_1$ &  $(1+z)^{\frac{3}{2}(1+w_0-\Omega_{\text m, 0} w_1)} [\Omega_{\text m, 0}(1+z)^3 + \Omega_{\text{de}, 0} ]^{\frac{w_1 - w_0}{2}}$
\end{tabular}%
\\
\caption{DE EoS used in our analysis and their corresponding dimensionless Hubble parameters calculated with Eq.~\eqref{eq:H}.}
\label{tab:EoS}
\end{table}

Our best fit values of the parameters are shown in Table \ref{tab:parameters} of Appendix \ref{ap:tables}. Note that in Eq.~\eqref{eq:H} the parameter $\Omega_{\text{de},0}$ and $ w_{\text{de}}$ appear just one time in the form of a product. Therefore, the parameters of conventional EoS parameterization will have a degeneracy with $\Omega_{\text{de},0}$ that cannot be disentangled with the parameter inference procedure (this degeneracy can be seen explicitly in the Hubble parameters of $w_0$, $w_{\text{CPL}}$ and $w_{\text{BA}}$ in Table \ref{tab:EoS}). For this reason, we gave the best fit values of the products $\Omega_{\text{de},0} w_0$ and $\Omega_{\text{de},0} w_1$ in Table \ref{tab:parameters}. This is not the case for the proposed EoS $w_\Lambda$, since it has a dependency on $\Omega_{\text{de},0}$ which breaks the degeneracy, as can be seen in its Hubble parameter and in our results given in Table \ref{tab:parameters}.

\begin{figure}[!htbp]
    \centering
    \includegraphics[width=9cm]{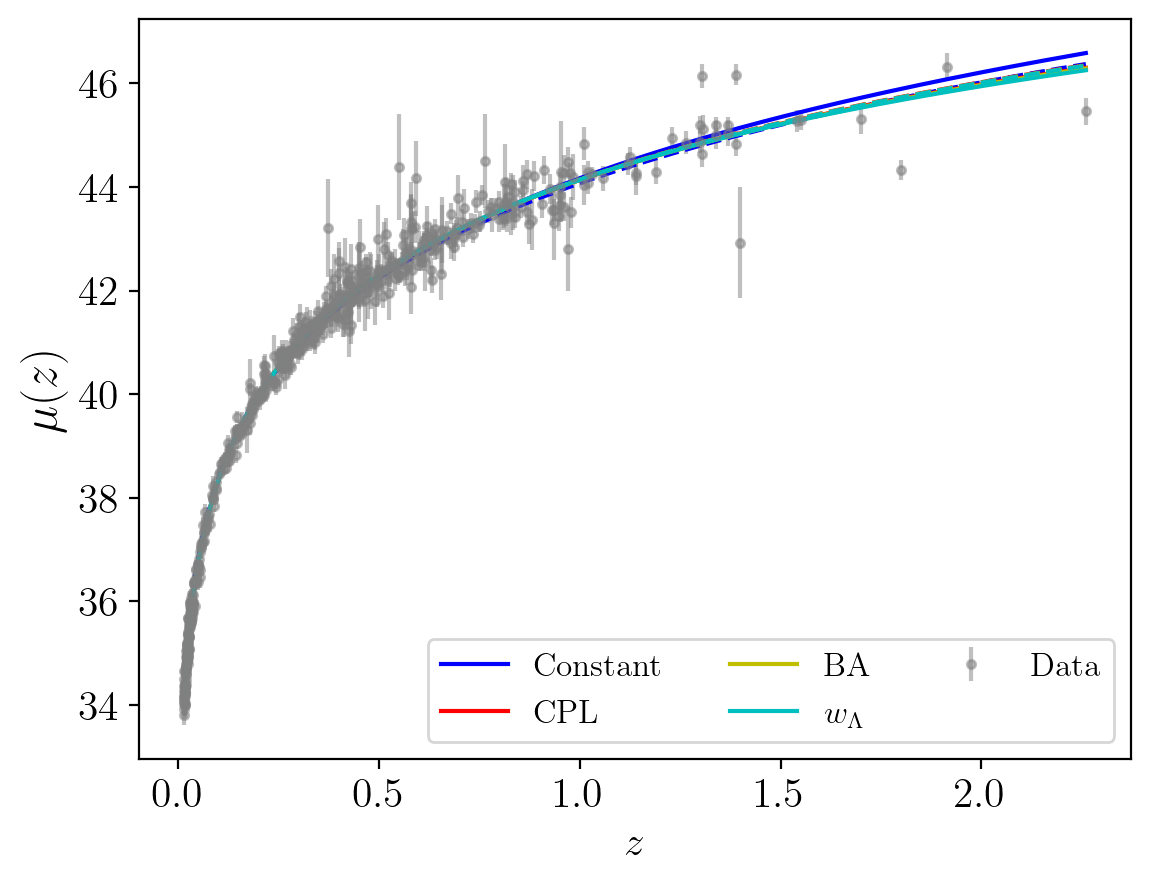}
    \caption{Supernovae data with Pantheon R19 calibration (gray points) and the best-fit curves of the modulus distance $\mu$ as a function of the redshift $z$ [Eq.~\eqref{eq:modulus_distance}] for the four parameterizations analyzed (see Table \ref{tab:EoS}), each color representing a different parameterization. The continuous (dashed) curves are calculated with a Bayesian analysis using the Pantheon (Pantheon+OHD) data.}
    \label{fig:mod_distance}
\end{figure}

\begin{figure}[!htbp]
    \centering
    \includegraphics[width=9cm]{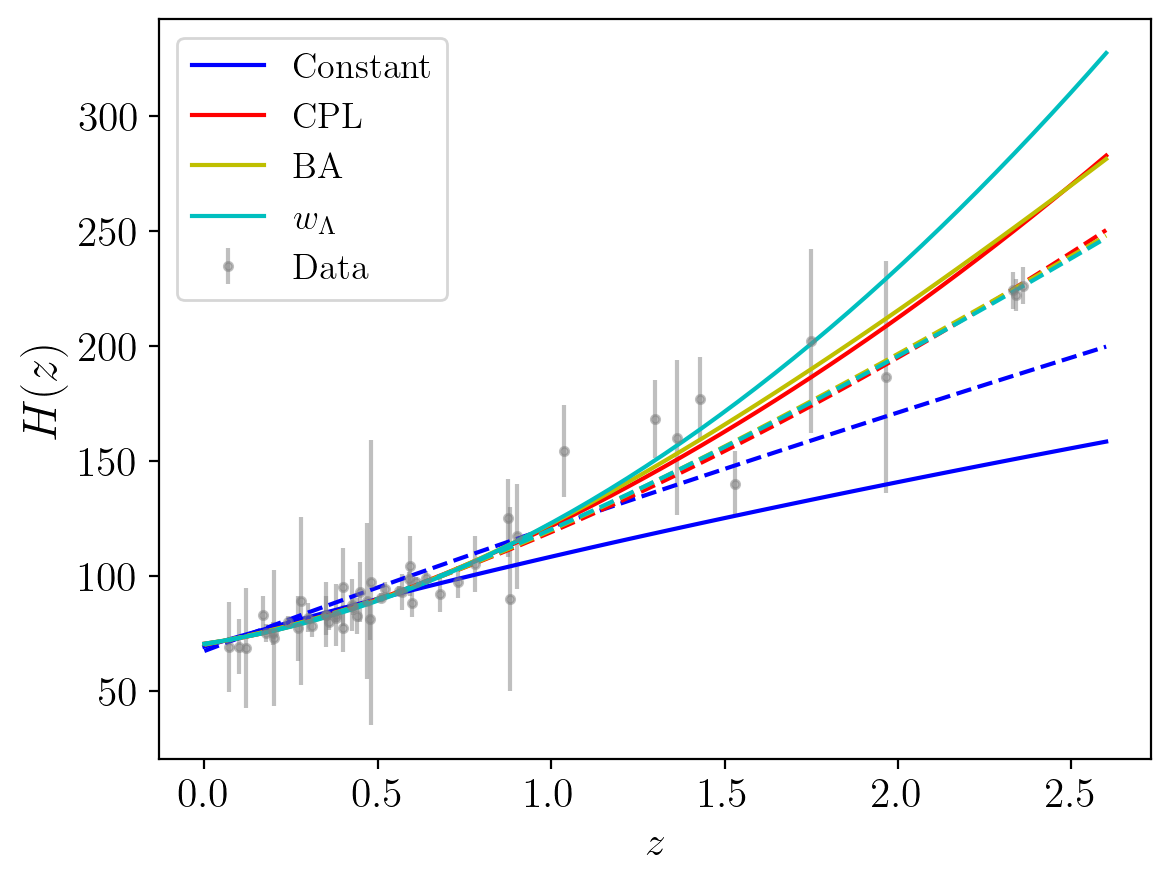}
    \caption{Observational Hubble data (gray points) and the best-fit curves of the Hubble parameter $H$ as a function of the redshift $z$ for the four parameterizations analyzed (see Table \ref{tab:EoS}), each color representing a different parameterization. The continuous (dashed) curves are calculated with a Bayesian analysis using the Pantheon (Pantheon+OHD) data.}
    \label{fig:CC}
\end{figure}

In Figures \ref{fig:mod_distance} and \ref{fig:CC} we show the modulus distance and the Hubble parameter for the four selected parameterizations, respectively. As we can notice, the evolution of the modulus distance of the four parameterizations follow the same trend of the supernovae data.
Nevertheless, in Figure \ref{fig:CC}  we can see that the constant parameterization yields a fitting curve that deviates significantly from the  trend of the OHD, therefore we will exclude this parameterization from next analyses.  On the other hand, the complexity of the other EoS produce qualitatively better fitting curves. Therefore, our analysis will only use the CPL, BA and $w_\Lambda$ parameterizations. 

\begin{figure}[!htbp]
    \centering
    \includegraphics[width=9cm]{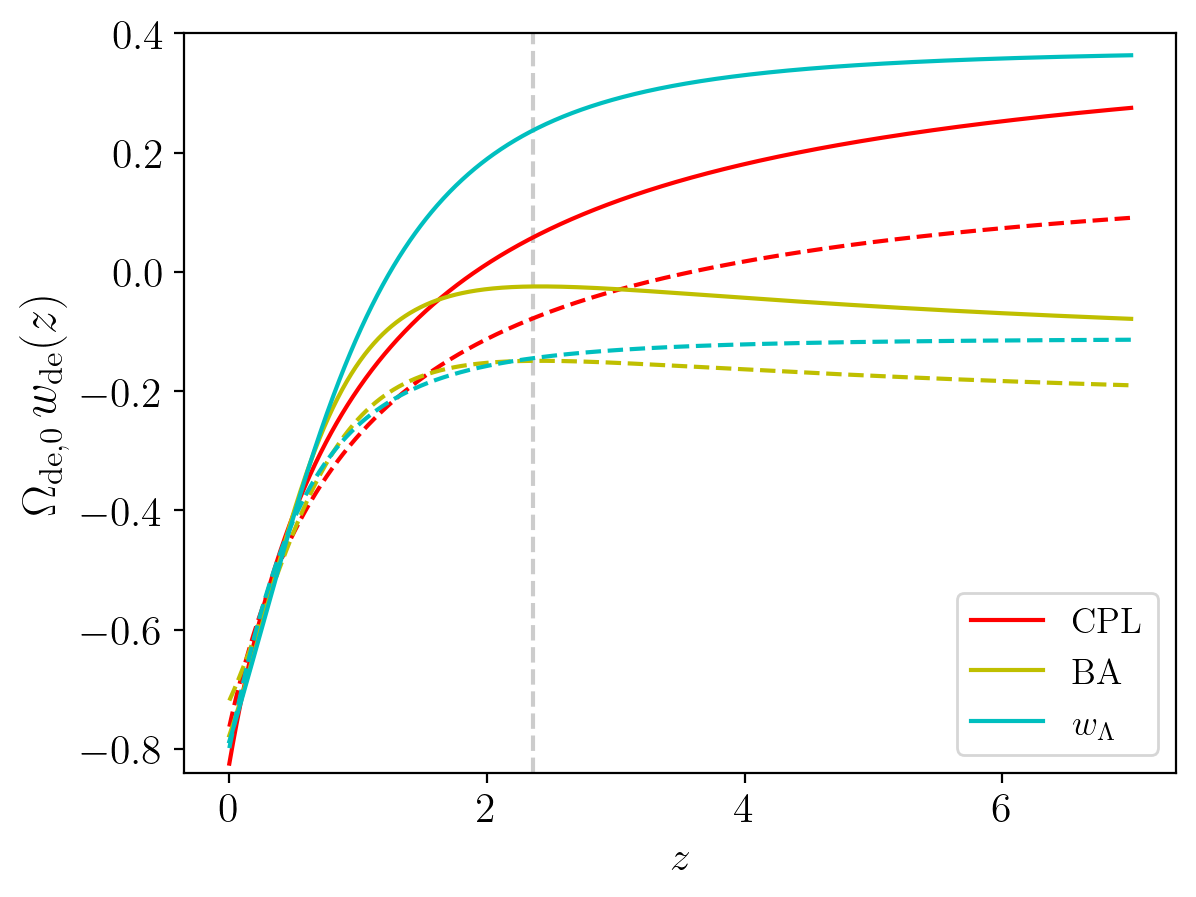}
    \caption{The product $\Omega_{\text{de},0} w_{\text{de}}$ as a function of the redshift $z$ for the four parameterizations analysed (see Table \ref{tab:EoS}), each color representing a different DE parameterization. The continuous (dashed) curves are calculated with the Pantheon (Pantheon+OHD) data. The vertical dashed line corresponds to $z=2.36$ which is the highest redshift in the data; the information beyond this line is an extrapolation of the observational data to show the expected behaviour of the DE EoS at higher redshifts in each model.}
    \label{fig:EoS}
\end{figure}

In Figure \ref{fig:EoS} we present the behaviour of  $w_{\text{CPL}}$, $w_{\text{BA}}$ and $w_\Lambda$ multiplied by $\Omega_{\text{de},0}$ (to deal with the degeneracy in the $w_{\text{CPL}}$ and $w_{\text{BA}}$ parameterizations). We notice that all the analyzed parameterizations allow an accelerated expansion of the Universe. However, the behaviour of $w_{\text{de}}$ changes allowing a positive value of $w_{\text{de}}$ for certain parameterizations at higher redshifts. Note that for the $w_{\Lambda}$ parameterization (where there is no degeneracy in its parameters) with the Pantheon$+$OHD data, we have $w_{\text{de}}(0) = -1.70 \pm 0.12$ (see Table \ref{tab:parameters}), which is relatively higher in comparison to the $\Lambda$CDM EoS $w_{\text{de}} = -1$. We expect a similar result in the CPL and BA parameterization (where there is a degeneracy): we have $\Omega_{\text{de},0} w_{\text{de}}(0) \sim -0.8$, but if $\Omega_{\text{de},0} \sim 0.7$, then $w_{\text{de}}(0)\lesssim -1$. Therefore, the considered DE EoS induce a \textit{delayed acceleration}, causing a smaller (or null) acceleration earlier on, and a faster acceleration at late periods, compared with the $\Lambda$CDM model. 

\begin{figure}[!htbp]
    \centering
    \includegraphics[width=9cm]{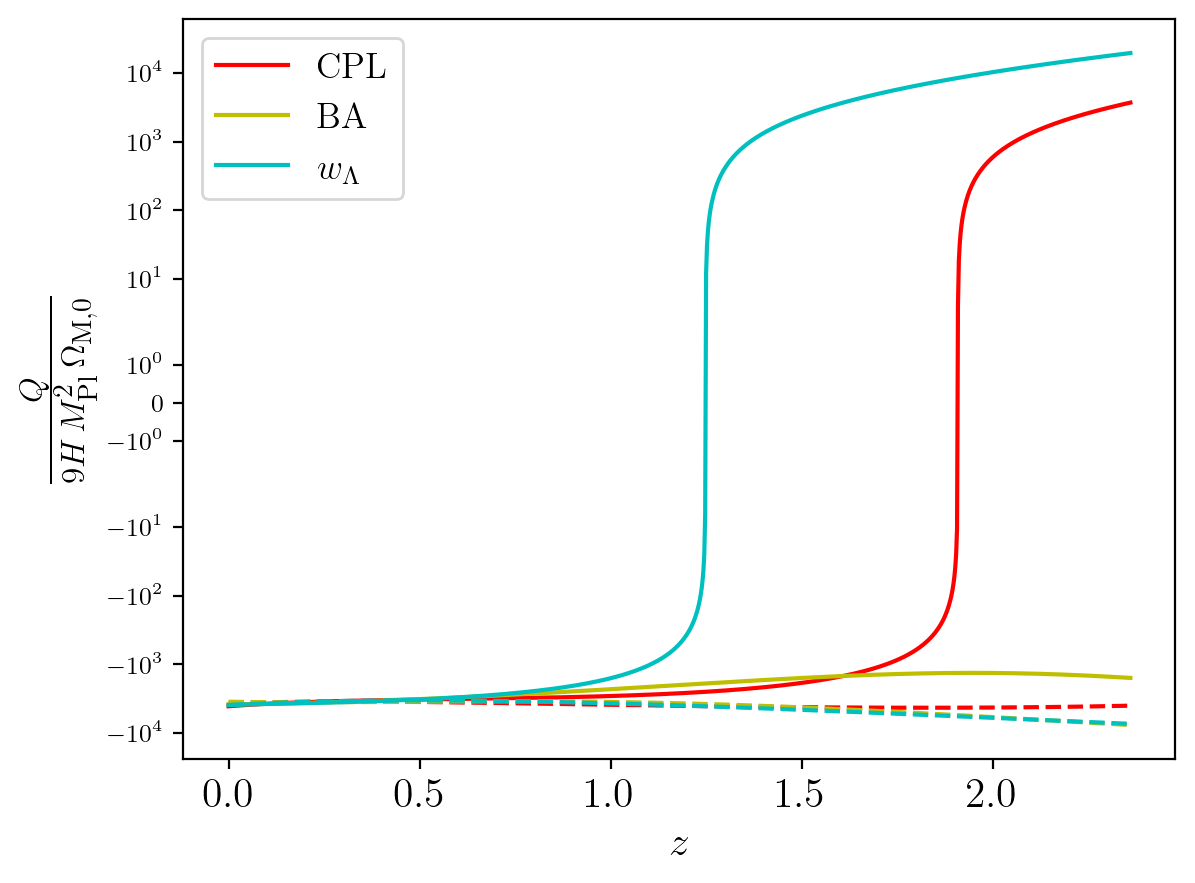}
    \caption{Rate of the interaction term $Q$ to the Hubble parameter $H$ as a function of the redshift $z$ [Eq.~\eqref{eq:interaction term dark dominated}] for the four parameterizations analysed (see Table \ref{tab:EoS}), each color representing a different parameterization: CPL (red), BA (yellow) and $w_{\Lambda}$ (blue). The continuous (dashed) curves are calculated with the Pantheon (Pantheon+OHD) data.}
    \label{fig:Q}
\end{figure}

Figure \ref{fig:Q} describes the rate behaviour $Q/H$ (multiplied by a constant). This term is a measure of the interaction compared with the rate of the expansion of the Universe. Since $Q/H <0$ today and $H>0$, then $Q<0$. Therefore, from Eq.~\eqref{eq:cont_eq rho_de}, we can conclude that DM is receiving energy from DE at these late stages of the Universe. Nevertheless, depending on the EoS, there may have been a time where the opposite transformation of energy happened ($Q>0$). Overall, the sign of $w_{\text{de}}$ determines the sign of $Q$, as can be seen from the plots and from Eq.~\eqref{eq:interaction term dark dominated}. 

\begin{figure}[!htbp]
    \centering
    \includegraphics[width=8cm]{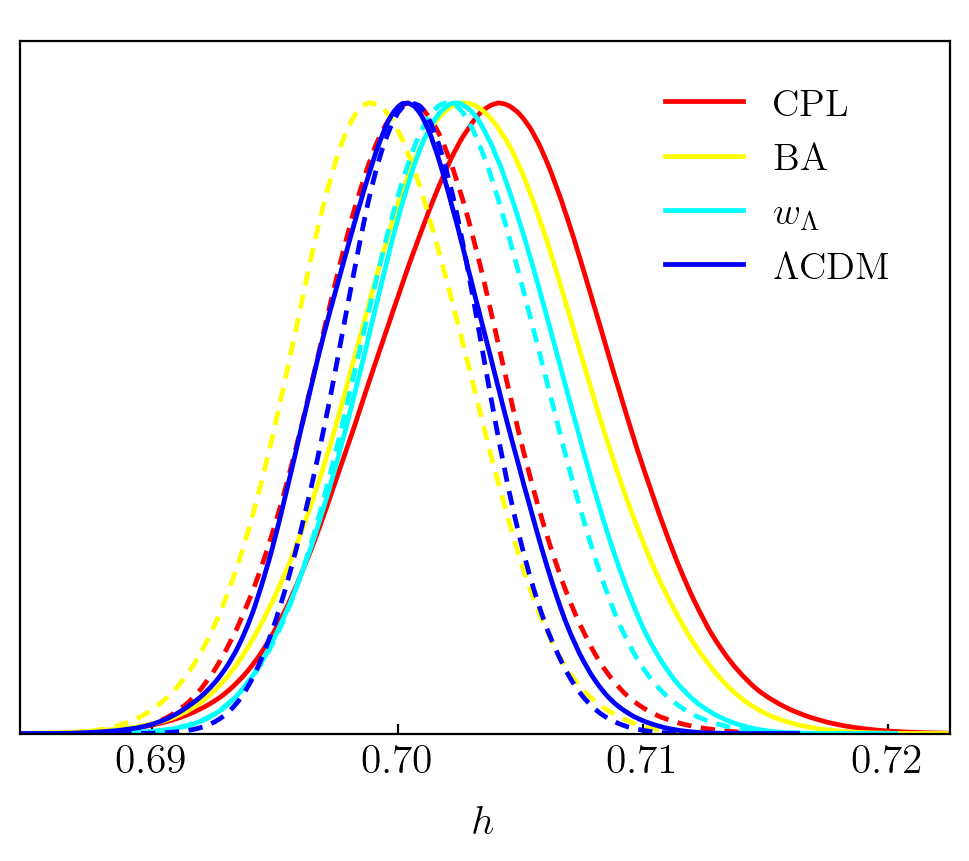}
    \caption{Posterior distributions of the reduced Hubble constant $h$ for the four parameterizations analysed (see Table \ref{tab:EoS}) and for the standard $\Lambda$CDM model. The continuous (dashed) curves are calculated with the Pantheon (Pantheon+OHD) data. 
    }
    \label{fig:h_posteriors}
\end{figure}

Finally, in Figure \ref{fig:h_posteriors} we show the marginalized distributions of the reduced Hubble constant $h$. For comparison, we have included the benchmark $\Lambda$CDM model \eqref{eq:H_LCDM} with $\Omega_{\text r,0} = 0$ and $k=0$. First, we notice that the inclusion of the OHD moves the posteriors towards smaller values of $h$ as a consequence of the functional form of the Hubble parameters obtained from the three parameterizations employed (see Table \ref{tab:EoS}): in order to fit the data in Figure \ref{fig:CC}, the right end of the continuous curves must be moved to produce the dashed lines, moreover, with the cosmic evolution, the left end (which determines $H_0$) is also affected. Other EoS parameterizations may introduce a different behaviour. Despite this shift towards smaller values of $h$, the CPL and $w_\Lambda$ parameterizations give a slightly higher value of $h$ with respect to the matter-DE dominated $\Lambda$CDM universe. Nevertheless, this change alone is not enough to give a value of $h$ consistent with late Universe observations, which generally estimate a value of $h \gtrsim 0.73$ \cite{Verde:2019ivm}. We expect that information about the DE EoS at higher redshifts (or, equivalently through Eq.~\eqref{eq:hubble_Qpar}, information about the interaction term at higher $z$) would be needed to increase the value of $h$ obtained from the statistical analysis of the parameterized IHDE model. Indeed, early Universe observations (high and median redshifts of the CMB and galaxy BAO data) were needed in the HDE model of \cite{h0_HDE} (with the future event horizon as the IR cutoff) to alleviate the Hubble tension with a delayed acceleration, which we expect is the same mechanism operating in this IHDE model (see Figure \ref{fig:EoS}). The aforementioned information about the DE EoS could be obtained, for example, with the cosmography approach which allows to reconstruct the kinematic evolution of the Universe acceleration at higher redshifts (up to $z\sim 10$ \cite{Escamilla-Rivera:2021vyw}) without assuming a cosmological model \cite{Escamilla-Rivera:2019aol, Capozziello:2013wha, Bolotin:2018xtq}.

\section{Conclusions}
\label{sec:conclusions}

In this paper we studied a new parameterized version of the IHDE model: a non-standard cosmological model which is based on the holographic principle and that allows an interaction between DE and DM. Our proposal uses a Hubble IR cutoff and a DE EoS redshift parameterization to quantify the interaction, instead of assuming it with a specific form. Our simple model \textit{solves the so-called coincidence problem of Cosmology}. 

Furthermore, using standard Bayesian inference procedures applied to observations coming from Pantheon, Cosmic Chronometers and Galaxy Clustering data sets, we showed that this simple model allows an accelerated expansion of the Universe in comparison to the results found in previous work where certain forms of the interaction were considered. Therefore, our work suggests that we can explain the late Universe observations with this simple parameterized version of the IHDE model.

Many specific features discussed in the analysis of our results depend on the chosen DE EoS parameterization. For this reason, we used the well-known CPL and Barboza-Alcaniz parameterizations. Nevertheless, with these choices we found a degeneracy in the Hubble parameter of this model that restricted our knowledge about some of the parameters involved. Therefore, we proposed a new DE EoS parameterization (motivated by the $\Lambda$CDM limit in this model) that allowed us to break the degeneracy in the parameters. More DE EoS of this kind could be used in future analyses.

We showed through plots the behaviour of the DE EoS, the interaction between DE and DM, and the posterior distribution functions of the reduced Hubble constant for the parameterizations considered. Depending on the redshift, these parameterizations allow an energy transfer from DE to DM, or vice versa. Furthermore, some of these parameterizations predict a slightly higher value of the Hubble constant, compared with the $\Lambda$CDM model. In fact, this model allows a faster accelerated expansion at later stages of the Universe, and a possibly smaller (or null) accelerated expansion at earlier times. This delayed acceleration (first discussed in \cite{h0_HDE}) may alleviate the Hubble tension. Nevertheless, to test that hypothesis in this model, we would need a specific form of the interaction term (or a DE EoS) valid at higher redshifts to perform the statistical analysis. The cosmography paradigm may give us the required information at higher redshifts, up to $z \sim 10$.

Natural suggestions for future work are the inclusion of more parameterizations of the DE EoS to keep studying this model in the low-redshift regime. Another option would be the use of a direct parameterization of the interaction term; the corresponding Hubble parameter could be calculated through Eq.~\eqref{eq:hubble_Qpar}. This redshift parameterization of the interaction may offer a more direct insight of the coupling between DE and DM. The next big step would be an analysis with higher redshifts (possibly through the cosmography approach) to study the Hubble tension in the IHDE model with a Hubble IR cutoff. This will require an interaction term that behaves qualitatively similar to those presented here, in order to fit the late Universe observations. Finally, the idea of parameterizing the DE EoS or the interaction term could be used with other IR cutoffs to maintain the agnostic position with respect to the interaction. 


\backmatter

\bmhead*{Acknowledgments}
CE-R is supported by PAPIIT UNAM Projects IA100220 and TA100122 and acknowledges the Royal Astronomical Society as FRAS 10147. This work is part of the Cosmostatistics National Group (\href{https://www.nucleares.unam.mx/CosmoNag/index.html}{CosmoNag}) project.
AG acknowledges financial support from ``Consejo Nacional de Ciencia y Tecnología'' (CONACyT) graduate grants program.


\begin{appendices}

\begin{sidewaystable}[!htbp]
\section{Best-fit parameters}
\label{ap:tables}
\vspace*{1cm}
\centering
\resizebox{19cm}{!}{%
\begin{tabular}{c|cc|cc|cc|cc}
 & \multicolumn{2}{c|}{Constant $w_0$} & \multicolumn{2}{c|}{$w_{\text{CPL}}$} & \multicolumn{2}{c|}{$w_{\text{BA}}$} & \multicolumn{2}{c}{$w_\Lambda$} \\ \hline
 & \multicolumn{1}{c|}{Pantheon} & Pantheon+OHD & \multicolumn{1}{c|}{Pantheon} & Pantheon+OHD & \multicolumn{1}{c|}{Pantheon} & Pantheon+OHD & \multicolumn{1}{c|}{Pantheon} & Pantheon+OHD \\ \hline
$h$ & \multicolumn{1}{c|}{$0.691\pm0.003$} & $0.672\pm0.003$ & \multicolumn{1}{c|}{$0.704\pm0.005$} & $0.701\pm0.004$ & \multicolumn{1}{c|}{$0.703\pm0.004$} & $0.699\pm0.004$ & \multicolumn{1}{c|}{$0.703\pm0.004$} & $0.702\pm0.004$ \\
$\Omega_{\text{de},0} w_0$ & \multicolumn{1}{c|}{$-0.569\pm0.021$} & $-0.433\pm0.011$ & \multicolumn{1}{c|}{$-0.825\pm0.066$} & $-0.763\pm0.028$ & \multicolumn{1}{c|}{$-0.778\pm0.055$} & $-0.719\pm0.024$ & \multicolumn{1}{c|}{---} & --- \\
$\Omega_{\text{de},0} w_1$ & \multicolumn{1}{c|}{---} & --- & \multicolumn{1}{c|}{$1.257\pm0.308$} & $0.976\pm0.073$ & \multicolumn{1}{c|}{$0.624\pm0.155$} & $0.472\pm0.035$ & \multicolumn{1}{c|}{---} & --- \\
$\Omega_{\text{m},0}$ & \multicolumn{1}{c|}{---} & --- & \multicolumn{1}{c|}{---} & --- & \multicolumn{1}{c|}{---} & --- & \multicolumn{1}{c|}{$0.203^{+0.028}_{-0.034}$} & $0.531^{+0.036}_{-0.030}$ \\
$w_0$ & \multicolumn{1}{c|}{---} & --- & \multicolumn{1}{c|}{---} & --- & \multicolumn{1}{c|}{---} & ---  & \multicolumn{1}{c|}{$-0.988^{+0.084}_{-0.034}$} & $-1.70\pm0.12$ \\
$w_1$ & \multicolumn{1}{c|}{---} & --- & \multicolumn{1}{c|}{---} &  --- & \multicolumn{1}{c|}{---} & --- & \multicolumn{1}{c|}{$0.470\pm0.058$} & $-0.237^{+0.072}_{-0.063}$
\end{tabular}%
}
\caption{Best-fit values for the four parameterizations considered (see Table~\ref{tab:EoS}), for both Pantheon and Pantheon+OHD data.}
\label{tab:parameters}
\end{sidewaystable}
\vspace*{\fill}

\end{appendices}


\bibliography{references}


\end{document}